# Ultrafast Control of Lifetime in High-Q Anisotropic Plasmon Polaritons

Romain Tirole[1*], Giacomo Venturi[1*], Emroz Khan[1], Lin Jing[1], Lin Nan[2], Nicola Melchioni[2], Andrea Mancini[2], Antonio Ambrosio[2] & Andrea Alù[1,3]†

[1] *Photonics Initiative, Advanced Science Research Center, City University of New York, New York, USA*

[2] *Center for Nano Science and Technology, Fondazione Istituto Italiano di Tecnologia, Milan, Italy*

[3] *Physics Program, Graduate Center, City University of New York, 365 5th Avenue, 10016, New York, USA*

* These authors contributed equally.

† aalu@gc.cuny.edu

## Abstract

Losses are a major roadblock in the technological implementation of surface plasmons at optical frequencies. The recent emergence of $MoOCl_2$, a correlated van-der-Waals material with strongly anisotropic optical properties, offers new avenues to circumvent this limit. We report the far-field observation of high-Q surface plasmon polaritons in this material, arising from the anisotropic hybridisation of surface plasmons and dielectric modes. We then explore nonlinear pumping of intraband electrons to the conduction band in these structures, leading, contrary to intuition, to an abrupt increase in the lifetime of the polariton resonance, despite the injection of hot electrons. This counterintuitive phenomenon stems from the competition between photon and plasmon excitations in a lenticular polariton resonance, yielding a largely tuneable lifetime at ultrafast speeds.

## Main

## Introduction

As quasiparticles arising from light-matter interactions, surface polaritons exhibit sharp optical resonances with large momenta beyond the light cone, highly sensitive to geometry and material properties. Among them, surface plasmon polaritons have served as the workhorse of the polaritonic field for the past two decades, and have found widespread applications in sensing[1] and photocatalysis[2]. At the same time, applications of plasmons have been often hampered by optical loss[3]. In particular, in the visible and UV frequency range, below 600 nm wavelength, classic metals, van-der-Waals materials and doped semiconductors/transparent oxides still fall short of supporting long-lived optical resonances with surface-bound field profiles[4].

In the nonlinear optical regime, the trade-off between local field confinement and losses in plasmonic materials is also pronounced. Because plasmons are electronic in nature, their optical response can be modulated efficiently at ultrafast timescales: the sharp change in time of the optical properties of plasmonic materials induced by all-optical pumping enables THz-speed applications such as optical[5] or magnetic[6] switching, holography[7], charge transfer for plasmonic photocatalysts[8], relativistic synthetic motion[9] and other time-varying media implementations[10]. Yet, all-optical pumping is often associated with an increase in material losses. The strong nonlinear optical properties of plasmonic materials rely mainly on resonant state absorption, which results in a systematic increase in the electronic temperature and thus the electron scattering rate at the origin of Ohmic losses in the medium. As a result, ultrafast optical pumping generally broadens plasmonic resonances and reduces their quality factor.

Recently, a layered van der Waals anisotropic metal - $MoOCl_2$ - has drawn attention due to its surprisingly long-lived hyperbolic plasmon polaritons,[11,12] despite the strong damping exhibited by its correlated electrons. The pronounced anisotropy and hyperbolicity of the material are due to Peierls distortions, causing the electronic conduction band to cross the Fermi level along the [010] axis ($x$), thereby making it metallic, while the other two [100] ($y$) and [001] ($z$) axes remain dielectric from visible to mid-infrared frequencies. Though this exotic behaviour is still under investigation by both experimentalists and condensed matter theorists[13–15], it is clear that $MoOCl_2$ presents new opportunities for the field of plasmonics. This potential is twofold: first from a fundamental physics approach, where the unique electronic behaviour of the material could break new barriers, and second from a wave-physics point of view, where the strong anisotropy of the crystal and its integration within nanostructures opens access to a diversity of polaritonic resonances [16–18].

Here, we exploit the natural anisotropy of $MoOCl_2$, and the unique coexistence of plasmonic-like and photonic-like modes in this material to observe long-lived hybrid surface modes at visible frequencies. As the two-in plane axes of the material exhibit metallic and dielectric properties, their respective plasmonic and waveguide modes can hybridise. We observe the self-hybridisation of these two modes through far-field momentum space experiments, by coupling light to a thin-film of $MoOCl_2$ with a metasurface. We report short-range surface plasmons with Q-factors above 40 in the visible range at wavelengths below 600 nm, and photonic modes with Q-factors of 80 at longer wavelengths. This performance places $MoOCl_2$ as a leading plasmonic material, on par with silver[19,20] or graphene in their respective frequency ranges[21,22].

Remarkably, we show that, by optically driving electrons in the metallic conduction band to higher energy states, the Q-factor of the material not only can be dynamically tuned, but also increased to higher values, with higher resonance lifetimes, despite the increase in losses resulting from a higher electron temperature in the conduction band. We attribute this unique effect to the complex dynamics of the self-hybridised polariton, building on the interplay between the dielectric and plasmonic nature of the medium and their mode competition within the grating/hybrid material metasurface.

## Polariton modes of $MoOCl_2$

Thanks to its large anisotropy, $MoOCl_2$ presents a wide variety of subwavelength polaritonic resonances across multiple wavelength ranges [11]. A common approach to characterise the electromagnetic modes of an anisotropic thin film is to visualise its isofrequency contours (IFCs), namely horizontal cuts of the dispersion relation at fixed wavelength, displaying the modes distribution in the ($k_x$,$k_y$) plane. IFCs can be computed numerically through the transfer matrix method (TMM) by evaluating the imaginary part of the reflection coefficient [23] (see Methods), as shown in Fig. 1a for a 100 nm-thick $MoOCl_2$ on a glass substrate for a free-space wavelength of 610 nm. In $MoOCl_2$, the hyperbolic plasmon polaritons visible at larger momenta have recently drawn attention[11], but we also notice a lenticular polaritonic mode (pink curve) hybridized with a photonic waveguide mode (green curve) for in-plane momenta at an angle with respect to the metallic and dielectric crystal axes of the material. Interestingly, these resonances are much sharper in this region of momentum-space, with IFCs similar to the predicted hybridisation between guided and leaky modes at the interface of an isotropic plasmonic material and a uniaxial dielectric cladding[24]. In the case of $MoOCl_2$, both modes originate from the material dispersion itself rather than an enforced external anisotropy, and the term self-hybridisation may be better suited to describe this crossing.

In a far-field measurement configuration, these subwavelength modes can be visualized in reflection or transmission measurements, when the in-plane momentum mismatch between free-space photons and polaritons is bridged by grating coupling[25]. A periodic structure can provide the momentum $\boldsymbol{G}$ to access modes below the light-cone, allowing us to image resonances within the numerical aperture of the objective centred about $\boldsymbol{G}$. To capture how the crossing of the plasmonic and photonic modes will present itself in experiment, we model the system with an added superstrate consisting of a 100 nm thick $Si_3N_4$ grating implemented as an effective medium (see Methods). A 10 nm protective layer of $Al_2O_3$ is also included in the simulation to reproduce the experiment. As can be observed in Fig. 1b, the modal dispersion of our metasurface strongly depends on the in-plane direction of momentum, as expected for a highly anisotropic material (and anticipated by Fig. 1a for bare $MoOCl_2$). Along the principal axes, the system supports two distinct uncoupled modes reflecting the in-plane optical responses of $MoOCl_2$: a plasmonic mode along the metallic $x$ axis (orange curve), and a waveguide-like mode along the dielectric $y$ axis (blue curve). However, approaching an in-plane angle $\theta \sim 45^{\circ}$, these modes interact and display an anti-crossing, revealing their hybridization into mixed plasmonic-photonic polaritons (purple curves). We aim to demonstrate the existence of these polaritons, understand their nature and explore their interplay in ultrafast experiments.

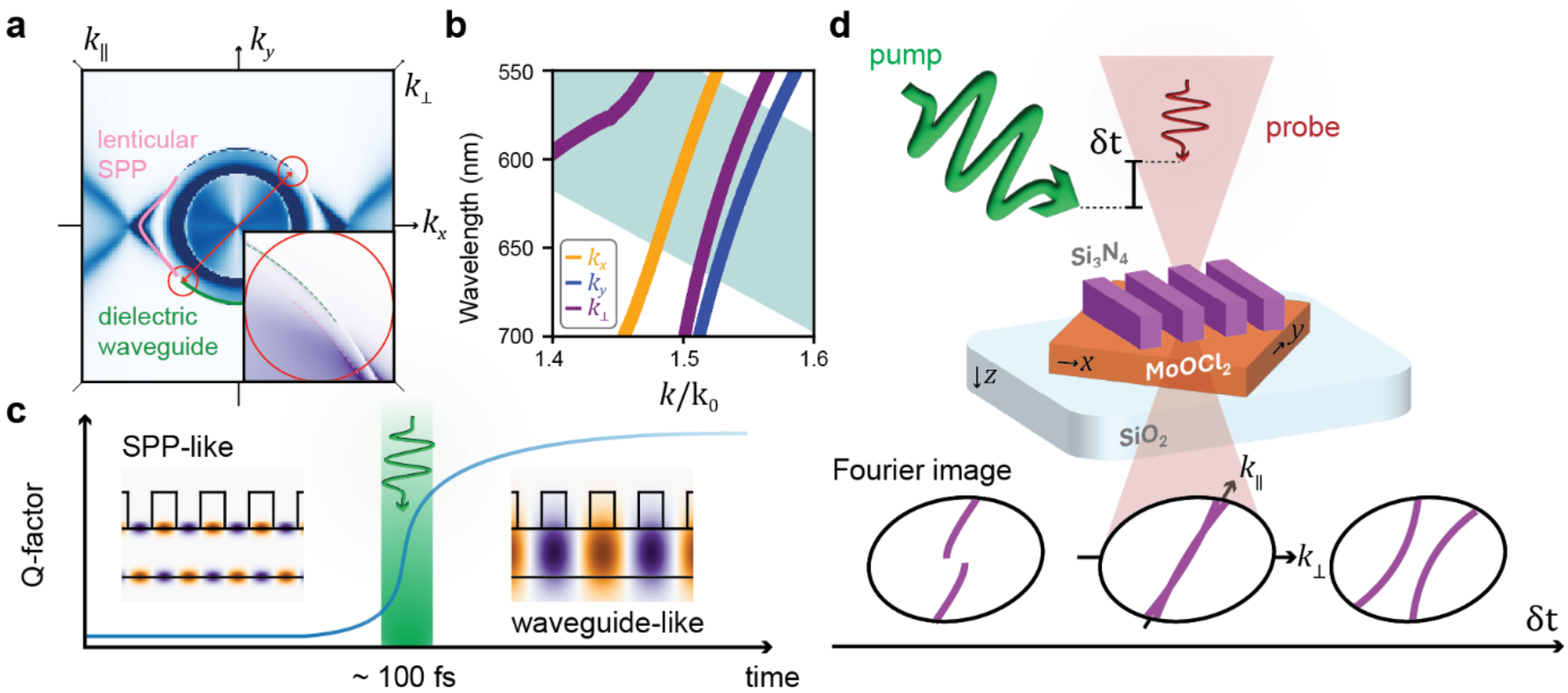

**Fig. 1: Plasmonic and waveguide photonic mode hybridisation in $MoOCl_2$. a** Simulated iso-frequency contour for a 100 nm thick $MoOCl_2$ flake at a wavelength of 610 nm, showing the dispersion of the lenticular surface plasmon polariton (pink curves) and the photonic mode (green curves). The red circles and arrows represent the momentum-space region observed through a 0.28 NA objective as in our experiments. The inset shows the computed transmission from TMM in the region corresponding to the red circles. **b** Calculated dispersion of the different modes in our metasurface: for a material with plasmonic dispersion along $x$ (orange curve) and a photonic dielectric waveguide dispersion along $y$ (blue curve), the crossing of the effective index waveguide at an intermediate angle of 45° leads to a splitting of the band (purple curves). The shaded area highlights the momenta covered by the objective numerical aperture. **c** Through ultrafast pumping of electrons in the conduction band, we control within an ultrafast timescale the character of these nanoscale modes in frequency and momentum space. **d** Diagram of pump-probe experiments in the back focal plane: in a transmission microscope, the probe (red arrow) is coupled to the subwavelength modes through a grating, and dips are recorded by imaging the back-focal plane of the transmission objective. These images are recorded as a function of delay between the pump (green arrow), inducing intraband absorption, and the probe (red arrow) experiencing the modulation of the material.

Nonlinear optical experiments are particularly motivated by the interplay between the plasmonic and photonic nature of these resonant modes, which could yield a powerful knob for all-optical control of light. This is an unprecedented opportunity to switch Q-factor of an optical device with high-contrast, and modulate the near-field distribution at will within an optical pulse's timescale, as sketched in Fig. 1c. To experimentally investigate this regime, we employ the pump-probe scheme depicted in Fig. 1d. Specifically, two non-degenerate 120 fs pump and probe pulses are focused onto the device using a 0.28 numerical aperture (NA) objective. The pump pulse wavelength is set to induce intraband absorption for carriers along the metallic axis of $MoOCl_2$, at a wavelength of 550 nm, while the probe interacts with the modulated plasmonic response at longer wavelengths. The back-focal plane of a second 0.28

NA transmission objective is then imaged onto a camera to measure the ($k_{\perp}$,$k_{\parallel}$) momentum-resolved transmission of the device as a function of pump-probe delay. ($k_{\perp}$,$k_{\parallel}$) are defined as the vector coordinates at an angle of 45° from ($k_x$,$k_y$) and centred at the momentum kick $\boldsymbol{G}$.

## Momentum-space observation of self-hybridisation

In order to address the hybrid mode from the far-field, a grating was designed to provide the required momentum to access those modes from the far-field (see Supplementary Information). An optical microscopy image of the device whose optical properties are studied in this work is presented in Fig. 2a. The grating was deposited and etched on top of the $MoOCl_2$ flake at an angle of 45 degrees to the metallic axis (see Methods). Scanning electron microscopy of a test flake confirms the successful fabrication of such nanostructures (see Supplementary Information). Our permittivity model of the flake, corresponding to that presented by Melchioni et al. [15], is shown in Fig. 2b. $MoOCl_2$ undergoes an epsilon-near-zero transition at 520 nm along the $x$ axis, exhibiting plasmonic nature at longer wavelengths, while the dielectric $y$ axis shows a dielectric behaviour within this range. The out-of-plane $z$ axis is also dielectric with little dispersion within this wavelength range. Though other models are reported in literature [11,21], those are all consistent within our wavelength range of interest and the excellent agreement between modelling and experiment in the present work justifies our choice of permittivity.

We start our experimental investigation by measuring the far-field transmission spectrum of the device, and observe the far-field coupling to these confined modes. The orange curve in Fig. 2c shows two clearly visible dips at 578 and 608 nm, confirming the efficient coupling to the two polaritonic modes enabled by the grating. As the polarization state of the plasmon polariton is a mixture of TE and TM, the absorption peak is visible for both $x$ and $y$-polarized light. We choose to use $y$-polarized light as a probe for experimental measurements to achieve higher optical contrast of the resonances and to prevent any photocarrier induced dynamics from the beam in pump-probe experiments. The transmission of the same flake outside of the grating coupling area (shaded orange curve) shows no such dip, demonstrating the absorption resonances originate from modes beyond the light cone.

For a better description of these confined modes and polaritonic dispersion, we perform Fourier-space imaging measurements to probe the momentum distribution of the excited modes in the nanostructure. The spatial frequencies of the transmitted probe beam are recorded using a monochrome CCD camera by imaging the back-focal plane of the transmission objective. To record the dispersion, the probe is made no longer broadband but nearly monochromatic, with a full-width at half-maximum (FWHM) of 3 nm, and swept across the wavelength range 550-700 nm (see Methods). Example momentum-space transmission images are shown in Fig. 2d-f for probe wavelengths of 595, 610 and 625 nm, with Rigorous Coupled Wave Analysis (RCWA, see Methods) simulated images showing good agreement (Fig. 2g-i). At lower wavelengths (Fig. 2d,g), the broader plasmonic mode dominates the response of the material, while the waveguide shows weak absorption. As the modes cross at longer wavelengths (Fig. 2e,h), the waveguide starts dominating the optical absorption. As the wavelength keeps increasing (Fig. 2f,i), the modes move towards the edge of the image and disappear due to momentum mismatch. Note that the symmetry of the images is due to the symmetric coupling of the grating to positive and negative momenta at the same time (as depicted by the red circles in Fig. 1a) [25].

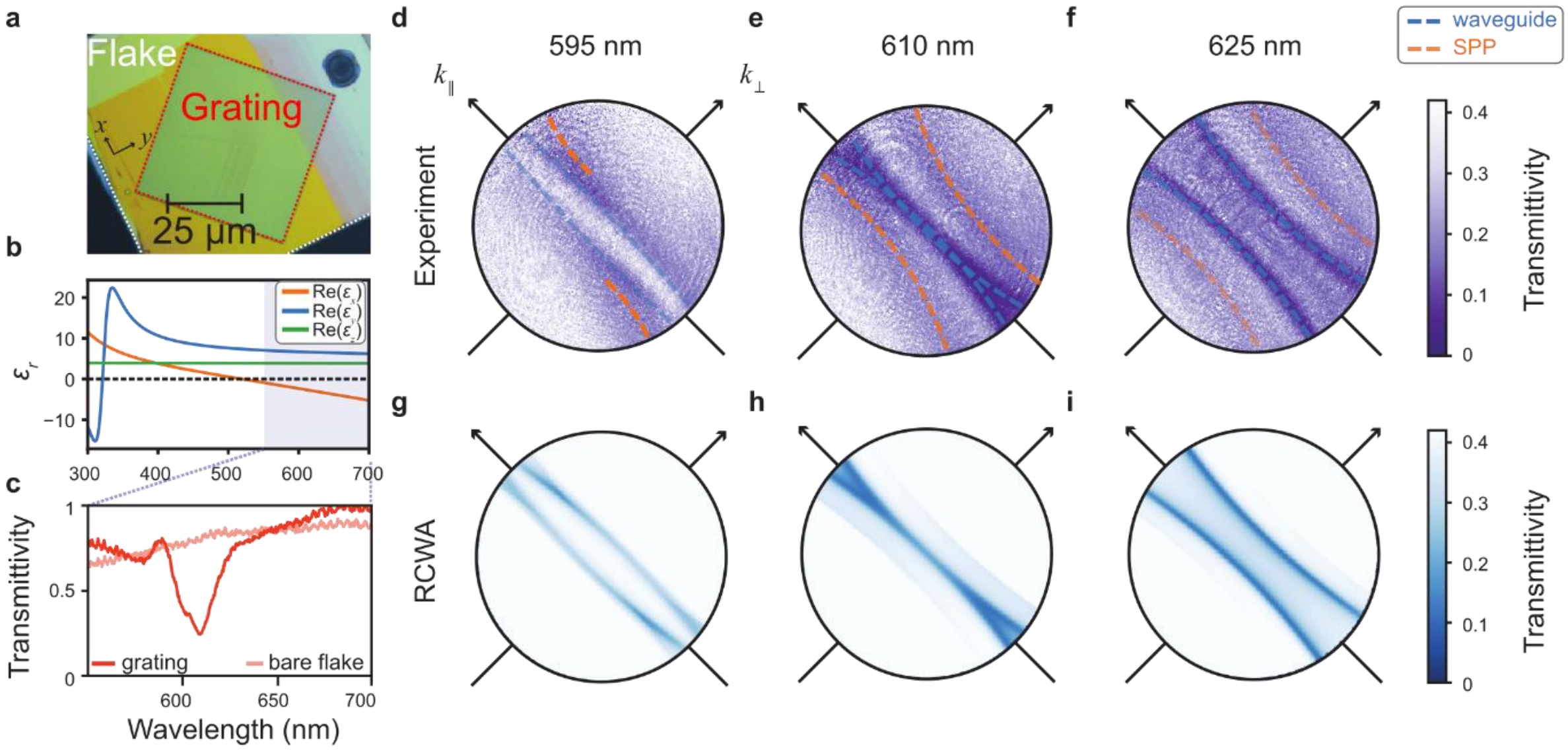

**Figure 2. Momentum-space imaging of the subwavelength modes. a** Optical microscopy image of the $MoOCl_2$ flake and fabricated gratings. **b** Dielectric permittivity of $MoOCl_2$ along its metallic [010] ($x$, orange), dielectric [100] ($y$, blue) and out-of-plane dielectric [001] ($z$, green) axes. **c** Measured transmittivity of a 100 nm thick $MoOCl_2$ flake with (continuous curve) and without (transparent curve) a grating, for $y$-polarized light. The coupling to the subwavelength modes is evident in the strong dip in transmission. **d-i** Experimental (d-f) and simulated (g-i) momentum-space transmission images of the device, for wavelengths of 595 nm (d,g), 610 nm (e,h) and 625 nm (f,i). The orange and blue dashed lines highlight the plasmonic and waveguide modes respectively.

While RCWA provides an efficient computational way to simulate momentum-space images, we use Finite-Difference Time-Domain (FDTD, see Methods) to target specific points of the dispersion and extract field distributions and Q-factors. The simulation results shown in Fig. 3a,b depict the out-of-plane field $E_z$ in the $(x_\perp, z)$ plane where $x_\perp$ is the spatial coordinate along the vector $\widehat{x_\perp} = \cos(\pi/4)\,\hat{x} + \sin(\pi/4)\,\hat{y}$, with the simulated probe having momentum along $k_\perp$ at the peak resonance. At 595 nm (Fig. 3a), the resulting electric field is surface-confined, with opposite phase on either side of the $MoOCl_2$ flake, demonstrating the short-range surface plasmon nature of the mode, in agreement with TMM field simulations from literature [11]. The field penetrates the medium and the substrate to some amount due to the hybridisation of the modes and the added radiation loss induced by the grating. On the other hand, the field at 650 nm (Fig. 3b) is confined within the volume of the flake, showing the waveguide nature of the mode at this wavelength.

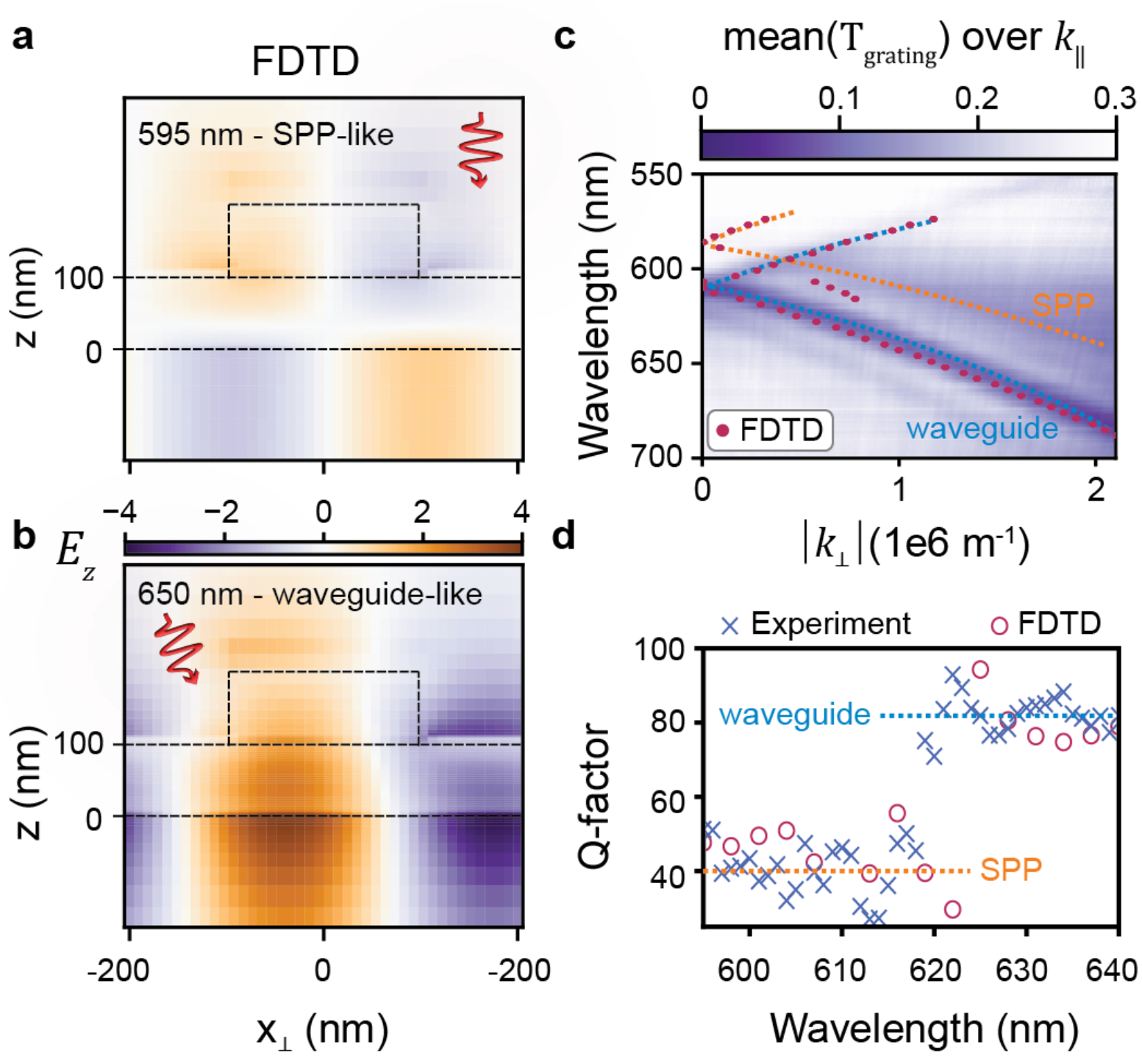

**Figure 3. Dispersion and mode characterisation of self-hybridised plasmons in $MoOCl_2$. a,b** Out-of-plane electric field distributions in the grating plane $(x_\perp, z)$, showing the plasmonic and photonic (volume-confined) character of the modes. The red arrows depict the angle of incidence of the probe to give the appropriate coupling momentum. **c** Experimentally measured transmission dispersion along the $k_\perp$ axis, constructed from images such as those shown in Fig. 2d-f. The numerically computed resonant wavelengths of the $MoOCl_2$ flake from FDTD (red dots) are superposed on the experimental data, while the guide dashed lines indicated the two hybridised modes – plasmonic (orange) and waveguide (blue). **d** Corresponding $Q$-factor of the main resonant branch, defined as $Q_k = k/\delta k$, experiment (blue crosses) against numerical simulations (red circles).

The simulation study of the near-field shows the sharp change in character of the modes with wavelength, a change that is strongly supported by the experimentally measured dispersion curve of the metasurface. We show in Fig. 3c the dispersion map defined as $T(\lambda, k_\perp) = \int T(\lambda, k_\perp, k_\parallel) \mathrm{d}k_\parallel$ where $T(\lambda, k_\perp, k_\parallel)$ is the measured momentum-space transmission as a function of wavelength (as in Fig. 2d-f). Note that the **vectors** $\boldsymbol{k}_\perp, \boldsymbol{k}_\parallel$ and their respective scalars represent here distance and direction from the grating momentum $\boldsymbol{\Lambda}$ i.e. from the centre point of the momentum-space image. FDTD simulations show an excellent agreement with the measured dispersion, with the extracted resonant wavelengths (red dots) closely overlapping with the absorption dips. The hybridised modes, plasmonic and waveguide, are respectively highlighted with the guide orange and blue dashed lines. As the image is integrated over $k_\parallel$, the plasmonic mode appears here a lot weaker and broader as it is more strongly located in specific regions of the Fourier image, and varies strongly in $k_\perp$ over different $k_\parallel$, while the waveguide mode is observable at a comparably stable $k_\perp$ throughout all values of $k_\parallel$ (see Fig. 2d-f). Qualitatively, the plasmonic mode 'dominates' the absorptive behaviour of the metasurface below wavelengths of 620 nm, while the waveguide mode takes over at longer wavelengths as it becomes sharper.

To further quantify this, we extract from our momentum space images the Q-factor of the dominant resonance along $k_\perp$, defined as $Q_k = k/\delta k$ where $k$ is the momentum of light and $\delta k$ the FWHM of the resonance (see Methods). At shorter wavelengths, where the mode has surface plasmon characteristics, the Q-factor is lower, hovering at a value of 40, and as the probe wavelength increases the Q-factor undergoes a transition to the higher value of 80. This can be well understood in terms of the transition from the lossier plasmonic resonance to the longer-lived photonic mode. Analysis of the resonances at momenta away from $k_\perp$ (i.e. away from the centre of the Fourier image) shows the respective modes have comparable quality factors there, and that the Q-factor is not strongly dependent on $k_\parallel$ throughout the numerical aperture of the objective (see Supplementary Information).

To translate $Q_k$ into the more common frequency-domain Q-factor, we model the plasmon dispersion and extract the phase and group velocity to compute $Q_\omega = \omega/\delta\omega = v_p Q_k / v_g$. For an example wavelength of 600 nm, where the resonance has plasmonic nature, we compute $v_p/v_g = 1.75$ which combined with our experimental measurements gives $Q_\omega \sim 70$, very close to the FDTD-computed value of $Q_\omega = 69.7$ (see Supplementary Information).

## Ultrafast control of lifetime in high-Q lenticular plasmon polaritons

Nonlinear optics offer a variety of mechanisms for ultrafast control of plasmon polaritons. Direct photocarrier excitation usually provides the highest optical contrast by inducing strong changes in the electronic conduction band, in comparison to nonparametric processes relying on virtual excitations of the medium [26]. Here, we propose to optically pump intraband transitions along the $x$ axis, thus changing the electron effective mass by pushing electrons to higher energy states. For example, this technique has been shown to provide high optical contrast and plasmonic dispersion modulation in Indium-Tin-Oxide [27,28]. For our pump-probe experiments, we set the 120 fs pump pulses polarised along $x$ at 550 nm wavelength. We then probe with a cross-polarized (along $y$) white-light pulse, filtering for wavelengths above 600 nm. Fig. 4a,b shows such pump-probe data, where the relative change in transmission spectrum $\Delta T/T$ is recorded as a function of pump-probe delay (negative delay meaning the probe reaches the sample before the pump).

In the absence of a metasurface, optical pumping of the flake slightly decreases the transmittivity of the medium, as we experimentally show in Fig. 4a. This is due to the excited non-thermal electron distribution along $x$ in the conduction band thermalizing through electron-electron scattering along all directions including $y$ and $z$ axes, thus inducing free-carrier absorption in an otherwise lossless refractive index. The onset of the modulation is pump-pulse limited, as absorption and carrier thermalization are processes typically faster than 100 fs [3]. In metals, relaxation of hot electrons is usually mediated by electron-phonon interaction with a timescale of 100s of fs to a few ps. Here, the relaxation time is much longer, on the scale nanoseconds, which suggests exotic electron-phonon interaction as observed in [29] and calls for a better understanding of the specific relaxation processes in $MoOCl_2$ in a later work.

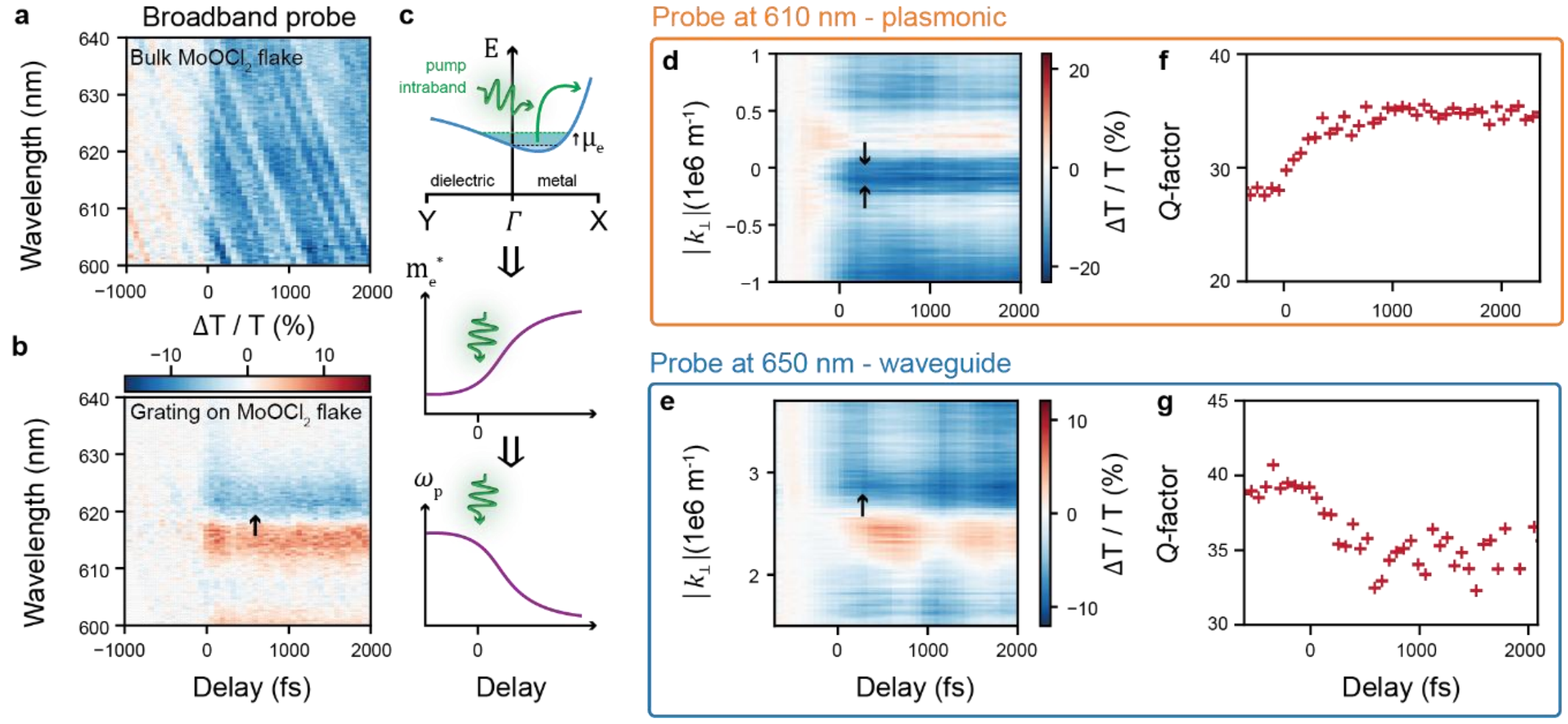

**Figure 4. All-optical ultrafast control of $MoOCl_2$ polaritons. a,b** Pump-probe spectroscopy of the device: using a broadband probe pulse, the evolution of the transmission as a function of delay is measured for both the bare flake at pump intensity of 215 GW/cm$^2$ (a) and the metasurface for a pump intensity of 138 GW/cm$^2$ (b). For the bare flake, an increase in losses reduces transmittivity, while for the metasurface the change in plasma frequency of the metallic axis shifts the resonance to longer wavelengths (black arrow). **c** This can be understood from the band structure of the material: the electrons along the $\Gamma - X$ are driven to higher energy states by the pump, thus increasing their average effective mass. This in turns reduces the plasma frequency, leading to a redshift in dispersion. **d, e** Momentum-space pump-probe measurement of the metasurface for narrowband probes at 610 nm (d) and 650 nm (e). At 610 nm, the push to larger momenta brings the resonance closer to the crossing at the centre of the momentum-space image, while at 650 nm the resonance is moved away from the centre. **f, g** Corresponding measured Q-factor as a function of delay: at 610 nm, (f) a clear increase is observed, while at 650 nm (g) the quality factor decreases under optical pumping.

In comparison, when pumping the metasurface, a clear redshift of the resonance is observed in Fig. 4b. A redshift of the mode dispersion is expected along both $x$ and $y$ directions. Along $y$, free carrier absorption lowers the refractive index of the medium, thus reducing the tight confinement of the photonic waveguide. Along $x$, by examining the band structure presented by Ruta et al. [12] and sketched in the top panel of Fig. 4c, one can expect the effective mass at higher electron energies to increase, thus leading to a decrease in the Drude plasma frequency and a redshift of the plasmonic dispersion. A common feature between these two mechanisms (Drude vs. dielectric dispersion) is the change in momentum of the mode dispersion: red shifting the resonance reduces the in-plane momentum of the mode. To validate our hypothesis, we reproduce this pump-probe experiment with a different grating with momentum along the plasmonic axis, thus probing only the metallic component of $MoOCl_2$ (see Supplementary Information). We find a similar red shift of the plasmonic resonance, of the order of 6 nm, with negligible broadening. This is compatible with a 2% decrease in the plasma frequency of $MoOCl_2$, with FDTD simulations reproducing this red shift both for the grating presented here and that in the Supplementary Information probing the plasmonic axis. On the other hand, losses seem to play a relatively negligible role in the optical pumping here, which could be attributed to the relatively small change of chemical potential within the conduction band.

To better quantify the effect of optical pumping on the confined polariton modes, we perform back focal plane pump-probe experiments to observe the evolution of the dispersion as a function of delay. The probe is switched from a broadband white light to a comparatively narrowband pulse. As the probe is 120 fs in duration, it exhibits a larger spectral width than the probe previously used in linear measurements: as a result, the resonances appear broader, with a lower $Q_k$.

To capitalise on the competition of the two resonances, we probe the evolution of the back-focal plane image at 610 nm, where the coupling to plasmon and waveguide-like modes is nearly overlapping due to the folding of the dispersion by the grating, with the plasmonic mode dominating the optical absorption. We observe a shift of the resonance towards the centre of the momentum-space image in Fig. 4d when pumping, thus making the resonance sharper. In comparison, when probing at 650 nm, where the plasmon has little role in the optical response of the device and the two resonances are clearly separated in momentum space, the absorption dip moves away from the centre of image in Fig. 4e.

This is consistent with our understanding of the material properties: the convexity of the modes below 610 nm shows they have momentum larger than the grating momentum $G$ (see Supplementary Information for a description of grating dispersion folding), which is expected at higher energies. When pumping, the decrease of the in-plane momentum will bring the resonances closer to the centre of our momentum-space image. On the other hand, for longer wavelengths (lower energies) whose modes have momenta lower than $G$, indicated by the opposite convexity, the resonance will move away from the center of the image.

From this data, we can extract the Q-factor of the metasurface as a function of pump-probe delay. Interestingly, as can be seen in Fig. 4f, $Q_k$ increases upon optical pumping for a probe at 610 nm. This is a counter-intuitive result as losses are known to increase in plasmonic systems under optical pumping, prohibiting increases in plasmonic lifetime. In the broadband far-field pump-probe measurement shown in Fig. 4b, the resonance's width stays roughly the same. This very weak increase in losses during intraband absorption is a first surprising feature, standing in stark contrast to other ultrafast plasmonic platforms [30,31].

The slight increase in losses is better observed when probing at 650 nm in Fig. 4g: here, the photonic waveguide mode, sensitive to the dielectric properties of the $y$ axis, observes a decrease in $Q_k$ from the free-carrier absorption. We thus hypothesise that the increase in Q-factor visible for lower wavelengths originates in the crossing and hybridisation of the plasmonic and photonic modes and their hybrid TE/TM nature: the surface plasmon is naturally more sensitive to the plasmonic dispersion of the material and is much more strongly affected by a change in plasma frequency or loss than the waveguide mode. The increase in Q-factor can then be understood through a Hopfield model of the coupling to the modes (see Methods). We write

$$Q_k \sim \frac{\beta_\pm}{\eta_{sp}^{(\pm)}\delta k_{sp} + \eta_{wg}^{(\pm)}\delta k_{wg} + \delta k_{rad}^{(\pm)}} \tag{1}$$

where $\eta_{sp}^{(\pm)}$ and $\eta_{wg}^{(\pm)}$ are the Hopfield coupling coefficients to the surface plasmon and waveguide-like modes respectively, and $\delta k_{sp}$, $\delta k_{wg}$ and $\delta k_{rad}$ are the broadening factors due to the intrinsic losses of the two modes and the radiative losses. Considering $\delta\beta_\pm/\beta_\pm \ll \delta Q_k/Q_k$, the ultrafast change in $Q_k$ must come from the denominator, i.e. the width of the observed resonance. Knowing both $\delta k_{sp}$ and $\delta k_{wg}$ see a negligible increase under pumping, we explain the increase in $Q_k$ by a strong change in $\eta_{sp}^{(\pm)}$ under optical pumping: the plasmonic mode undergoes a stronger change in its TE/TM distribution than the waveguide mode (see Supplementary Information), and in turn the coupling $\eta_{sp}^{(\pm)}$ to the resonance drops. As a result, the bandwidth of the absorption line goes from being dominated by $\eta_{sp}^{(\pm)}\delta k_{sp}(pump\ off)$ to being dominated by $\eta_{wg}^{(\pm)}\delta k_{wg}(pump\ on)$. This is confirmed by the matching in Q-factor between the modulated waveguide resonance at 650 nm and the switched hybridized resonance at 610 nm. In the end, the dispersion of $MoOCl_2$ is naturally engineered to enable the ultrafast switching between plasmon-like and waveguide-like resonances, thus increasing the interaction time of light with the system.

## Outlook

In conclusion, we have experimentally measured and rigorously modelled the hybridisation between polaritonic and photonic modes in a $MoOCl_2$ thin film. Remarkably, the resulting resonance exhibits high Q-factors accessible from far-field, regardless of the intrinsically short-range surface plasmon nature of the mode, and opening the opportunity to increase at ultrafast speeds the lifetime of light within the metasurface using all-optical pumping. This is a particularly unexpected finding: despite the injection of hot electrons in the system, the energy is stored in the medium for a longer time thanks to the hybridisation of the modes. By leveraging mode-engineering and optimal nanostructures[32–35], the high quality factor and the interplay between the different resonances of $MoOCl_2$ could bring nanoscale sensing or catalysis to a new level, or open a pathway to applications such as Q-boosting where the time-varying evolution of the metasurface breaks the time-bandwidth limit and enables indefinite energy storage[36–38]. Other time-varying applications may be explored: cyclic modulation of the $Q$ factor could amplify plasmons and counteract their losses[39], while a time-interface could excite magnetostatic waves or perform efficient frequency conversion of the light in the medium [40]. Our experiment also raises the question about the evolution of the electric field at the surface and in the bulk of the $MoOCl_2$ layer during the 100 fs scale switching, prompting for a further understanding of this non-adiabatic modal change of the device. This calls for a further theoretical and experimental push to achieve ultrafast control of subwavelength light confinement.

## Methods

Sample fabrication

The fabrication of the $Si_3N_4$ grating on $MoOCl_2$ relies mainly on a top-down etching process. The detailed process flow is as follows:

First, Au markers were patterned on a pre-cleaned fused silica substrate to facilitate the positioning of flakes that would be transferred later. The $MoOCl_2$ flakes were mechanically exfoliated from bulk crystals onto polydimethylsiloxane (PDMS) sheets and then transferred onto the substrate using an all-dry transfer method under an optical microscope. Subsequently, flakes of suitable dimensions were selected, and their thicknesses were measured using SNOM AFM. Based on these thickness measurements, the grating structure was simulated and optimized using the optical constants determined by ellipsometry (Woollam V-VASE) of the fabricated $Si_3N_4$.

Next, the grating fabrication began with $Si_3N_4$ deposition via plasma-enhanced chemical vapor deposition (PECVD, Oxford Plasma Pro System 100). The deposition thickness was closely monitored using a thin-film analyzer (Filmetrics F20). A reactive ion etching (RIE, Oxford Plasma Pro NPG80) step was then employed to fine-tune the $Si_3N_4$ thickness, ensuring a final $Si_3N_4$ layer of 100 nm. The substrate was subsequently coated with polymethyl methacrylate (PMMA) resist, and the metasurface pattern was defined using electron-beam lithography (EBL, Elionix 100 keV), during which the designed grating structure was aligned to the specific flake using the Au markers as reference. To mitigate charging effects during lithography, a thin Au (~10 nm) anti-charging layer was sputtered onto the PMMA. After lithography, the Au layer was removed by gold etchant (Transene, TFA), and the resist was developed using methyl isobutyl ketone (MIBK).

To form a positive grating structure, a thin $Al_2O_3$ hard mask (~15 nm) was deposited via electron-beam evaporation (AJA Orion 8E Evaporator System). This was followed by immersion in Remover PG and a lift-off process to transfer the pattern from the PMMA to the $Al_2O_3$ hard mask. Finally, the process concluded with a second RIE step ($CF_4$ 20 sccm, 100 W, 30 mTorr, 20 °C) to transfer the grating pattern from the $Al_2O_3$ mask to the underlying $Si_3N_4$ layer. A reference sample was used to calibrate the etched thickness, ensuring complete removal of the exposed $Si_3N_4$ while minimizing the exposure of the underlying $MoOCl_2$ to the $CF_4$ gas atmosphere, as the material proved reactive under these conditions. Notably, simulations indicated that the remaining $Al_2O_3$ on top had a negligible influence on the final optical performance.

Experimental measurements

Linear and nonlinear optical experiments are performed in a home-built microscope in transmission configuration with 10X, 0.28 NA apochromatic Mitutoyo objectives for both illumination and collection. The input beam polarization is controlled by a linear polarizer and a broadband half-wave plate before the illumination objective. The transmitted light is sent either to a monochrome Zelux CS165MU camera (for back-focal plane imaging) or to an OceanOptics Flame spectrometer (for transmission spectra). Imaging of the sample is done in reflection with a chromatic Zelux CS165CU camera. For linear measurements, a SuperK Fanium (NKT Photonics) delivers broadband ps pulses at a 78 MHz repetition rate, which are either sent directly to the microscope for transmission measurements or …. For back-focal-plane imaging, the broadband laser pulse is sent to a SuperK Select filter (NKT Photonics), tuning the pulse's wavelength between 550 nm and 700 nm with a ~3 nm bandwidth. For nonlinear measurements, an Astrella Laser (Coherent) sends 120 fs pulses to two Opera and Prime optical parametric amplifiers (Light Conversion), providing two independently tuneable light sources for pump-probe experiments. The pump is spectrally filtered out with long-pass filter in transmission. For broadband pump-probe measurements, white light is generated using a 3 mm-thick Sapphire crystal. In all measurements, the power of the probe is kept at a low level to avoid self-modulation..

Numerical modelling

In all numerical simulations, we employed the permittivity model presented in Melchioni et al. [15] , which assumes a diagonal and biaxial permittivity tensor. Despite $MoOCl_2$ being a monoclinic crystal, hence with non-zero off diagonal permittivity tensor elements, off-diagonal terms are assumed to be negligible, and a diagonal tensor is sufficient to model the experimental data (in line with previous studies [11,12]).

Transfer Matrix Method

The dispersion of polariton modes can be formally extracted by solving for the poles of the imaginary part of the reflectivity coefficient (usually p-polarized for isotropic materials). A standard approach in the polariton community is to numerically compute the reflection coefficient through a TMM, which has turned efficient in reproducing experimental data across different techniques from near-field to far-field data. TMM being an analytical method, it provides the exact solution to the Maxwell equations. We exploit an open source 4x4 transfer matrix formalism to compute the reflectivity matrix of a layered stack of homogeneous media with arbitrary anisotropic dielectric permittivity tensor as a function of in-plane momentum [23]. Since $MoOCl_2$ is a biaxial material, the eigenstates of electromagnetic modes are neither p- nor s-polarized. Hence, to retrieve all the modes in the system, we sum all the (imaginary parts of the) elements of the 2x2 reflectivity matrix $(r_{pp}, r_{sp}, r_{ss}, r_{ps})$ to keep track of the peaks corresponding to mixed polarization states.

The 2D plot of Fig. 1a is computed for a 100 nm-thick $MoOCl_2$ on a $SiO_2$ substrate. We fix the wavelength to 610 nm and plot the reflectivity for different in-plane rotation angles of the $MoOCl_2$ permittivity tensor (which is rotated around the z axis) to mimic momenta spanning the entire $(k_x, k_y)$ plane, thereby reconstructing the full reciprocal space. For Fig.1b, we fix the orientation of the permittivity tensor of $MoOCl_2$ and extract the peaks of the imaginary part of the total reflectivity as a function of wavelength. In this case, the goal is to inspect the modes of our metasurface, however since TMM can account only for homogeneous films, we employ an effective permittivity model for the 100 nm-thick grating layer (we averaged the $Si_3N_4$ and air permittivity, since the grating duty cycle is set to 50%), which is able to capture the physics of our device (see Supplementary Information).

Rigorous Coupled Wave Analysis (RCWA)

To simulate what to expect from the back-focal plane imaging measurements (Fig. 2), we employ an open-source RCWA code (RETICOLO) [41] and model our device as a layered stack made of: air, a 100 nm thick $Si_3N_4$ grating, 10 nm of $Al_2O_3$, 100 nm $MoOCl_2$, and $SiO_2$. We compute the diffraction efficiencies of the transmitted TE and TM modes for an illumination of the device from the grating side (as in the experiment). The numerical aperture of the objective is taken into account by feeding the system with in-plane momenta spanning all the angles within the illumination cone. The final images are obtained by projecting our input polarization - along the $MoOCl_2$ dielectric axis ($y$) - onto the TE and TM modes and weighting their diffraction efficiencies accordingly. For a proper account of the polarization mixing due to the objective focusing, we employed the Richards and Wolf formalism of the angular spectrum representation [42].

Finite Difference Time Domain (FDTD)

FDTD simulations were performed using the commercial Lumerical FDTD software. The grating was modelled with Bloch boundary conditions in the ($x$,$y$) plane, with perfectly matching layers along the $z$ direction. The grating orientation was kept along $x$ while the $MoOCl_2$ layer's permittivity was rotated by 45 degrees to emulate the experimental device. A plane wave source is then swept across angles within the experimental aperture of our objectives. For temporal Q-factor extraction, the electric field is monitored in time at the surface of the material (in air, 2 nm above the $Al_2O_3$ layer). For comparison with experiment (Fig 3c,d), a similar algorithm as for the experimental data is used.

Experimental Q-factor extraction

For each back-focal-plane image, a peak-finding algorithm locates the main absorption resonance. The FWHM of the resonance is then fitted along a series of angles from $k_\perp$ to account for the curvature of the band, with the highest value corresponding to the angle normal to the resonance curvature being retained. We then perform a weighted average over a 1e6 $m^{-1}$ range of values along $k_\parallel$. Manual inspection allows to check for consistency of the $Q$ factor at momenta different from $k_\perp$.

Hopfield model

We consider the basis of the plasmonic and waveguide modes, at fixed frequency $\omega_0$: $|sp\rangle$ and $|wg\rangle$, and the relative complex propagation constants $\tilde{\beta}_{sp}$ and $\tilde{\beta}_{wg}$ defined as:

$$\tilde{\beta}_j = \beta_j - i\frac{\delta k_j}{2}$$

$\delta k_j$ being the linewidth of mode $j$, as defined in the paper. We can write the non-Hermitian Hamiltonian, with losses taken into account:

$$H_{eff} = \begin{pmatrix} \tilde{\beta}_{sp} & g \\ g & \tilde{\beta}_{wg} \end{pmatrix}$$

Here, $g$ is the internal coupling between plasmon and waveguide mode. First, ignoring losses so that $\tilde{\beta}_j = \beta_j$, we can define the bare detuning in momentum as

$$\Delta = \beta_{sp} - \beta_{wg}$$

With this definition, the eigenvalues of $H_{eff}$ can be written as:

$$\beta_\pm = \frac{\beta_{sp} + \beta_{wg}}{2} \pm \sqrt{\Delta^2 + 4g^2}$$

We can derive the normalized eigenvectors as:

$$|\Psi_\pm\rangle = \frac{1}{\sqrt{1 + \left(\frac{g}{\beta_\pm - \beta_{sp}}\right)^2}} \left( |sp\rangle + \frac{\beta_\pm - \beta_{sp}}{2} |wg\rangle \right)$$

and the mixing angle as

$$\tan(2\vartheta) = \frac{2g}{\Delta}$$

so that the eigenvectors can be written as

$$|\Psi_+\rangle = \cos(\vartheta)\,|sp\rangle + \sin(\vartheta)\,|wg\rangle$$

$$|\Psi_-\rangle = -\sin(\vartheta)|sp\rangle + \cos(\vartheta)\,|wg\rangle$$

For each eigenvector, the relative weight of each original mode is called the Hopfield coefficient. These can be expressed as:

$$\eta_{sp}^{\pm} = \frac{1}{2}\left[1 \pm \frac{\Delta}{\sqrt{\Delta^2 + 4g^2}}\right]$$

$$\eta_{wg}^{\pm} = \frac{1}{2}\left[1 \mp \frac{\Delta}{\sqrt{\Delta^2 + 4g^2}}\right]$$

We can now insert back the losses in a perturbative manner, considering that the linewidth is small with respect to the energy of the mode (valid for the Q-factors considered here).

$$H_{eff} = H_0 + V_{loss} = H_0 - \frac{i}{2}\begin{pmatrix} \delta k_{sp} & 0 \\ 0 & \delta k_{wg} \end{pmatrix}$$

For sufficiently low damping, we can use first-order perturbation theory and write

$$\delta\tilde{\beta}_\pm = \langle \Psi_\pm | V_{loss} | \Psi_\pm \rangle = -\frac{i}{2}\left[\eta_{sp}^{(\pm)} \delta k_{sp} + \eta_{wg}^{(\pm)} \delta k_{wg}\right]$$

This gives the broadening from the two hybridizing modes. The radiative losses can be included as an additional loss channel by summing $\delta k_{rad}$ to the equation. We thus derive Eq. 1 of the paper as

$$Q_k^{(\pm)} = \frac{\Re(\tilde{\beta}_\pm)}{2\Im(\tilde{\beta}_\pm)} = \frac{\beta_\pm}{\delta k_\pm} = \frac{\beta_\pm}{\eta_{sp}^{(\pm)} \delta k_{sp} + \eta_{wg}^{(\pm)} \delta k_{wg} + \delta k_{rad}^{(\pm)}}$$

## Data availability

The data is available upon request.

## Funding and acknowledgments

Andrea Alù acknowledges support from the Simons Foundation and the Office of Naval Research. Antonio Ambrosio acknowledges funding from the European Union (ERC-2025-POC 2Dchiral N.101248056).

The authors thank Phillipe Lalanne for insightful discussions.

## Author information

These authors contributed equally: Romain Tirole and Giacomo Venturi.

Authors and affiliations:

**Photonics Initiative, Advanced Science Research Center, City University of New York, New York, USA**

Romain Tirole, Giacomo Venturi, Emroz Khan, Lin Jing and Andrea Alù.

**Center for Nano Science and Technology, Fondazione Istituto Italiano di Tecnologia, Milan, Italy**

Lin Nan, Nicola Melchioni, Andrea Mancini and Antonio Ambrosio.

**Physics Program, Graduate Center, City University of New York, New York, USA**

Andrea Alù.

Contributions:

R.T., G.V., A.A. and A.A. conceptualised the project. R.T. and G.V. performed experimental measurements. E.K. designed the coupling gratings. L.J., L.N., N.M. and A.M. fabricated the samples. G.V. led RCWA and TMM simulations. R.T. implemented FDTD simulations, nonlinear modelling and data analysis methods. A.A. and A.A. supervised the project and took on funding acquisition and project administration. R.T. and G.V. wrote the original draft. All authors contributed to the revision and discussion of the manuscript.

## Ethics declarations

### *Competing interests*

The authors declare no competing interests.

# Ultrafast Control of Lifetime in High Q Anisotropic Plasmon Polaritons

# Supplementary Information

Romain Tirole[1*], Giacomo Venturi[1*], Emroz Khan[1], Lin Jing[1], Lin Nan[2], Nicola Melchioni[2], Andrea Mancini[2], Antonio Ambrosio[2] & Andrea Alù[1,3]†

[1] *Photonics Initiative, Advanced Science Research Center, City University of New York, New York, USA*

[2] *Center for Nano Science and Technology, Fondazione Istituto Italiano di Tecnologia, Milan, Italy*

[3] *Physics Program, Graduate Center, City University of New York, 365 5th Avenue, 10016, New York, USA*

* These authors contributed equally.

† aalu@gc.cuny.edu

## Experimental setup

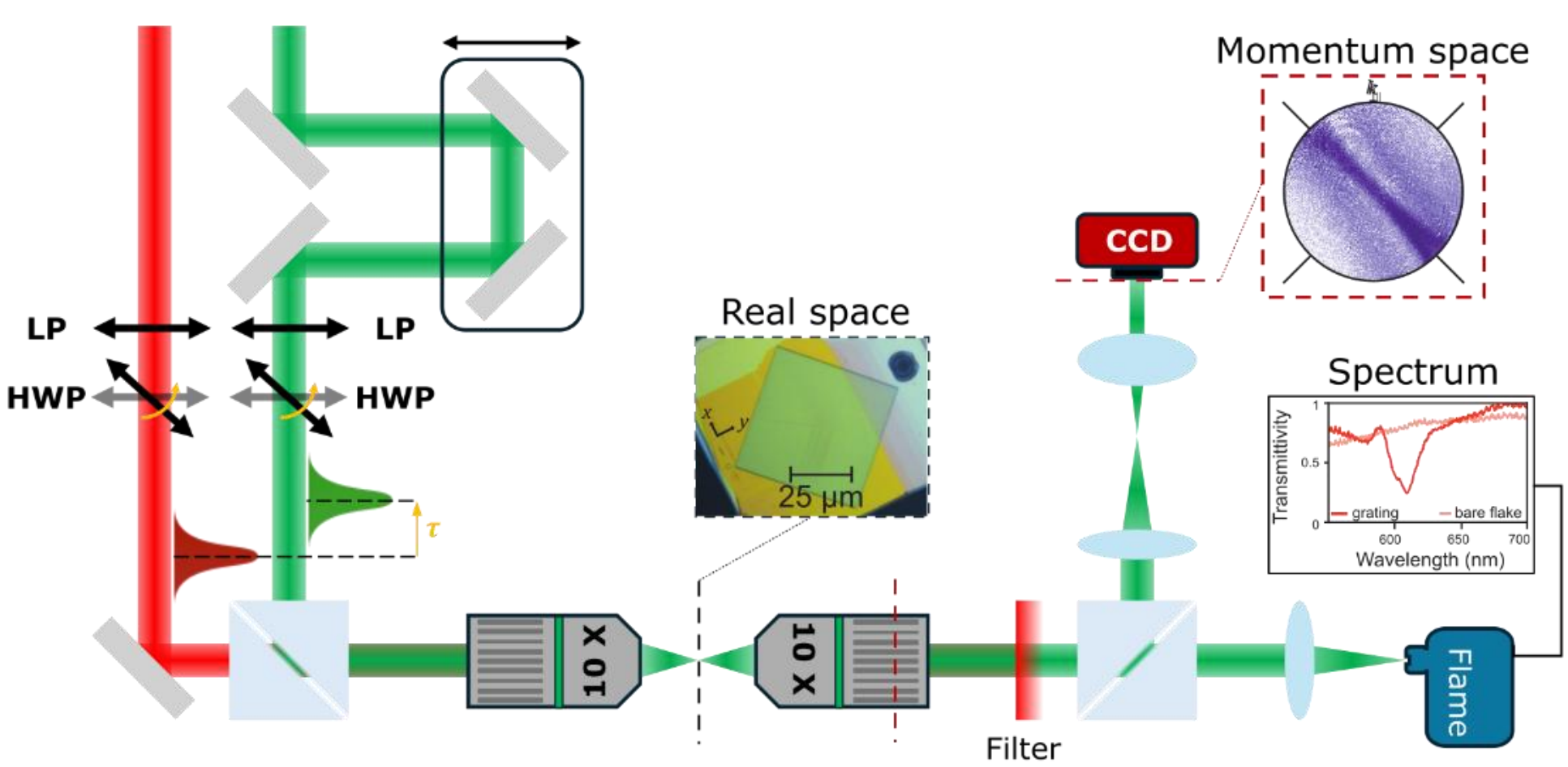

**Supplementary Figure 1.** Diagram of the experimental setup. The pump is represented by the red beam line, and the probe by the green beam line. The probe originates either from 3 nm wide long pulses for momentum-space reconstruction, or from 100 fs 'broadband' pulses for pump-probe experiments.

## Grating design

The grating are designed through parameter sweeping in finite elements simulations (Comsol Multiphysics). For a given flake thickness, corresponding to those measured on a variety of flakes in Atomic Force Microscopy, we vary the pitch $\Lambda$ and angle $\phi$ of the grating with regards to the $MoOCl_2$ flake's axes and illuminate the metasurface at normal incidence to emulate a momentum space image $T(k_x.k_y)$ where $k_x = \Lambda\cos\phi$ and $k_y = \Lambda\sin\phi$ dependent on grating parameters. The optimisation procedure aims setting the grating's final momentum (white dot in Supp. Fig. 2) at the centre of the modes' evolution.

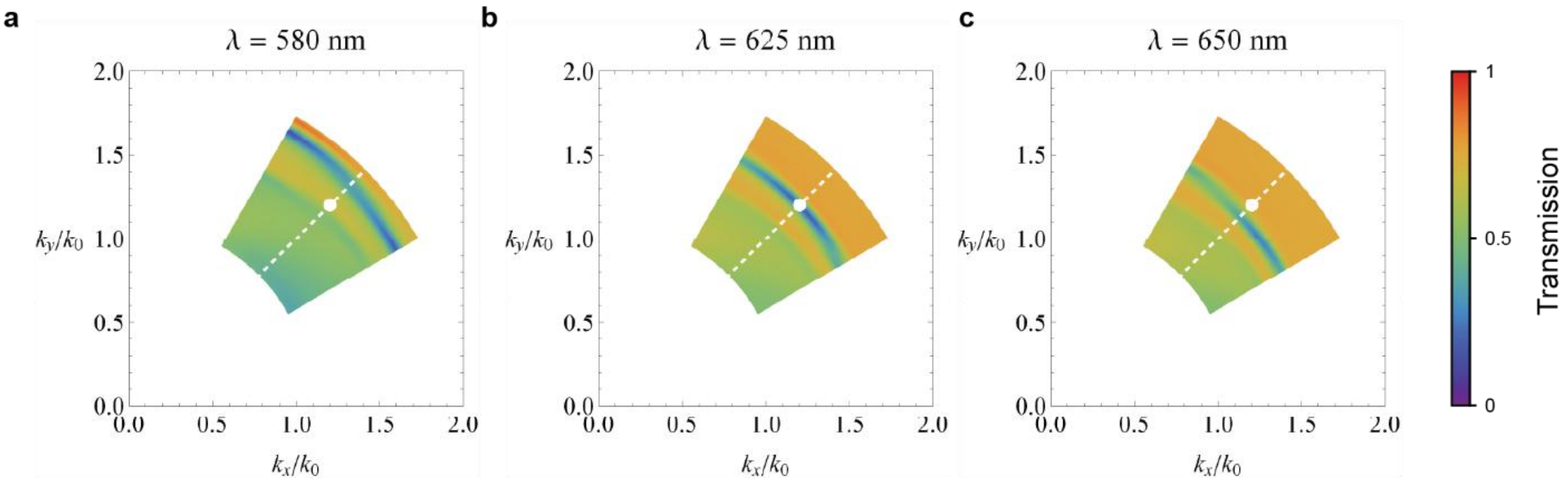

**Supplementary Figure 2.** Grating design: the pitch and angle of the grating are sweeped and transmission is recorded at normal incidence for various wavelengths: **a** 580 nm, **b** 625 nm and **c** 650 nm. The white dashed line highlights the grating angle while the white dot represents the momentum of the final grating design.

## Grating dispersion folding

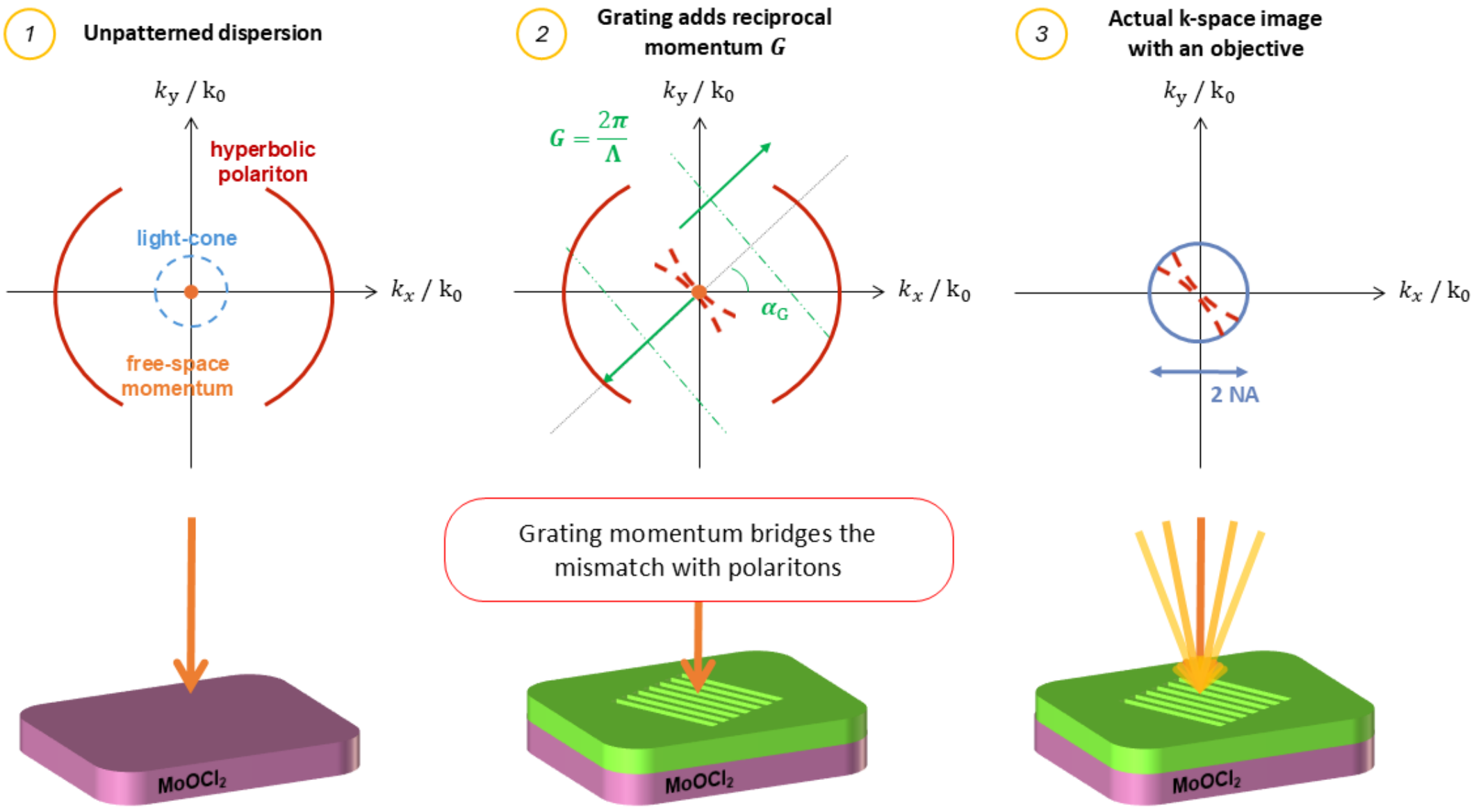

**Supplementary Figure 3.** Dispersion folding mechanism by the grating. Without a grating, polaritons below the light cone are not accessible from free space. Adding a grating folds the dispersion by a along symmetry planes defined by the grating momentum $\boldsymbol{G}$. This results in the observation of the resonances with inverted symmetry within the numerical aperture of a momentum-space imaging setup.

## Scanning electron microscopy

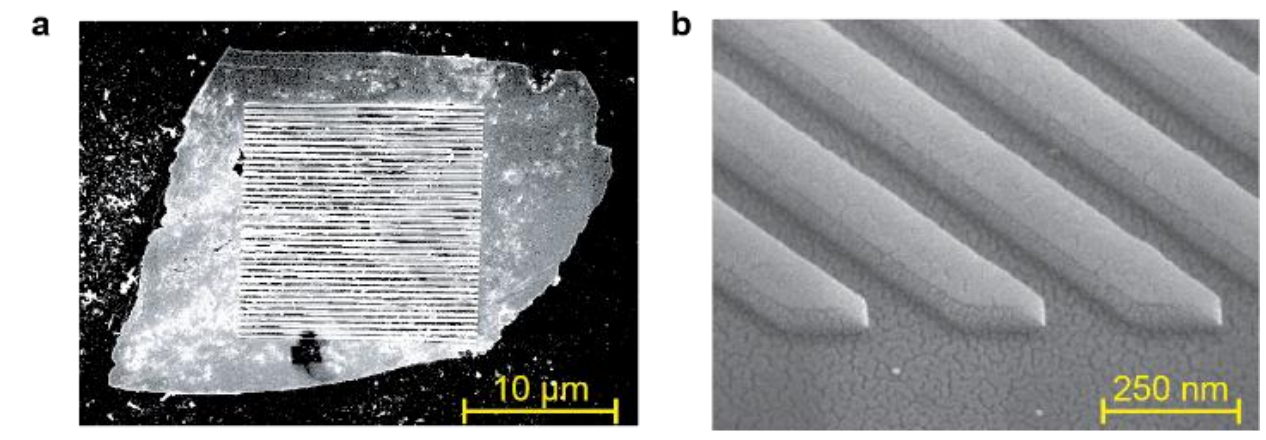

**Supplementary Figure 4.** Scanning electron microscopy image of test metasurfaces. **a** First fabrication test of a hybrid $MoOCl_2$/grating metasurface. The residue found around the grating is due to the strong reaction of $MoOCl_2$ with $CF_4$ during Reactive Ion Etching, effect mitigated in the following nanofabrication iteration (see Methods). **b** High-quality image of a test grating fabrication on a bare $SiO_2$ substrate. The surface roughness originates from the thin Au layer deposited for imaging.

## Q factor fitting

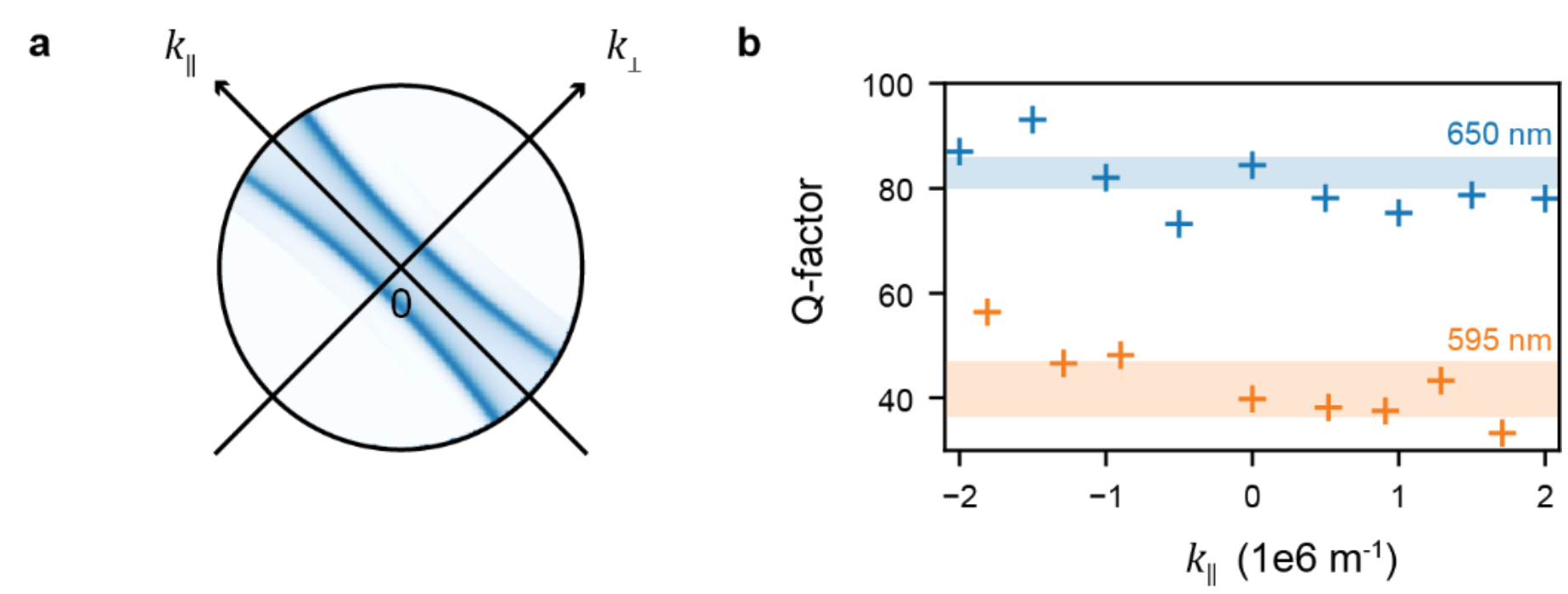

**Supplementary figure 5.** Q-factor extraction along $k_\parallel$. **a** RCWA simulation from Fig. 2**i**, with the $k_\perp$ and $k_\parallel$ directions illustrated. The algorithm finds the minima along $k_\perp$, and fits a Q-factor along different directions and extracts the minimal one to account for the curvature of the mode. The Q-factor is then averaged over a slice of 0.1e6 m$^{-1}$ along $k_\parallel$ along a central value (0 m$^{-1}$ for Fig. 3d). **b** Extracted values of $Q_k$ along various values of $k_\parallel$ for the two modes at different wavelengths.

## Pump-probe measurements

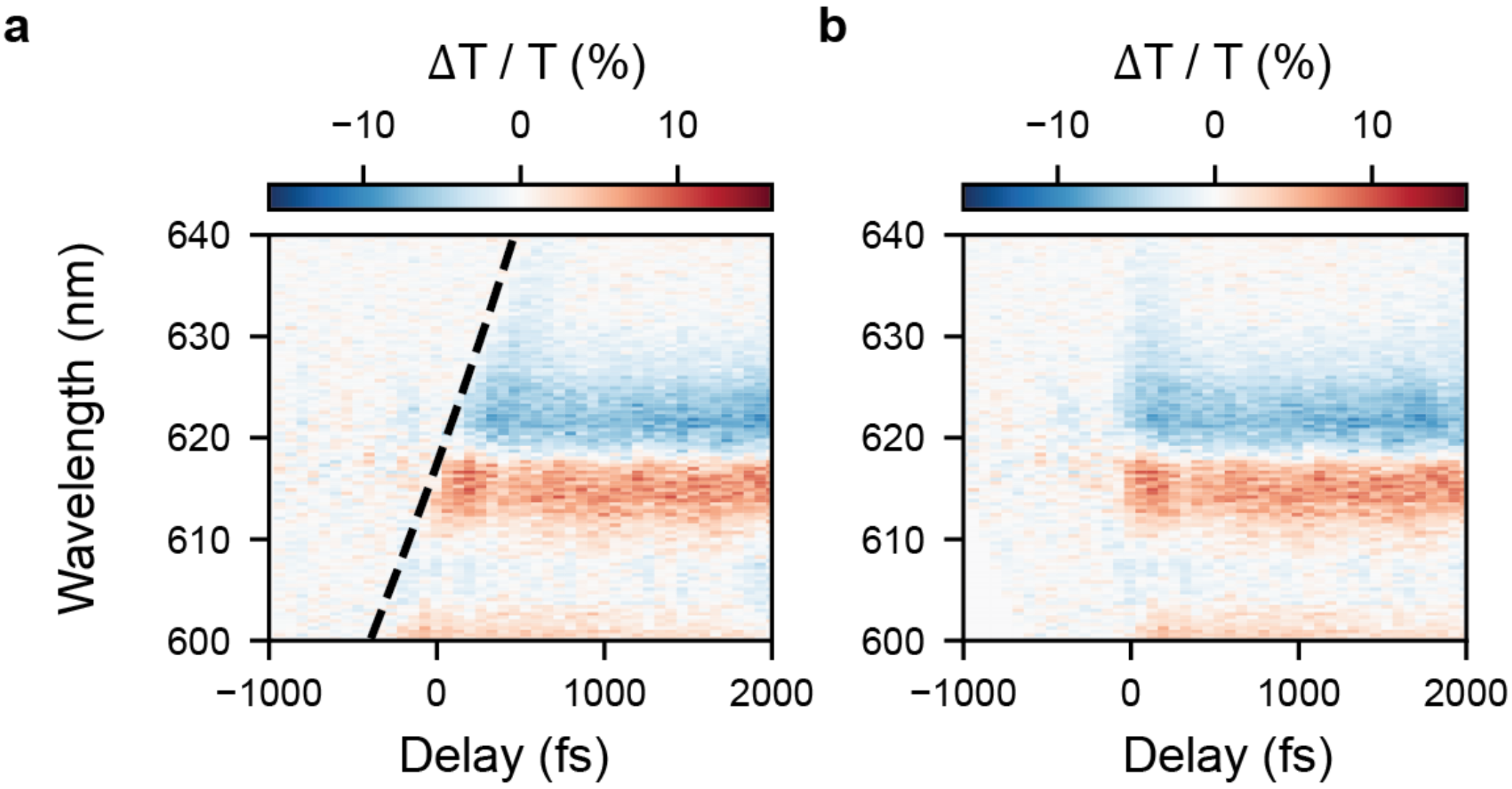

**Supplementary Figure 6.** Chirp correction in broadband pump-probe measurements. **a** Initial ultrafast spectroscopy data: the signal $\Delta T/T$ shows a different time zero for various wavelengths due to a strong

chirp induced during the generation of the white light in a 3 mm Sapphire crystal. The dark dashed line show the time zero is fitted as $t_0(\lambda) = c_0 + c_1(\lambda - \lambda_{\text{ref}}) + c_2(\lambda - \lambda_{\text{ref}})^2$. **b** Corrected pump-probe data, with the delay axis rescaled.

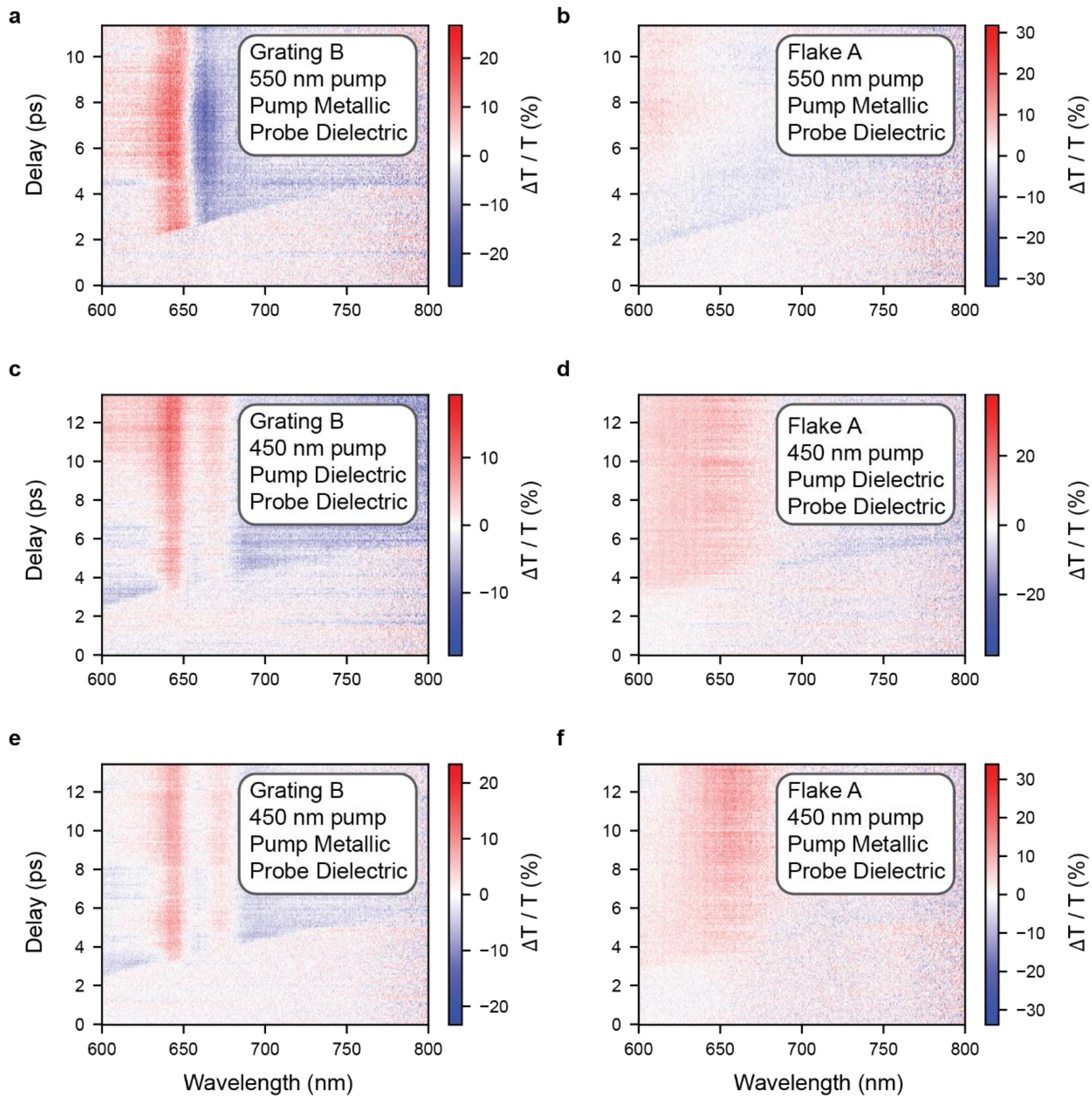

**Supplementary Figure 7.** Broadband pump-probe measurements. **a,c,e** Measured change in transmission of a grating with momentum along the plasmonic axis on a flake with 82 nm thickness and a plasmonic resonance at 650 nm for various pump wavelengths and polarisations. The measured shift in panel a allows us to estimate the change in plasma frequency of the material. **b,d,f** Measured change in transmission of the bare flake next to the grating measured in the main text for various pump wavelengths and polarisations.

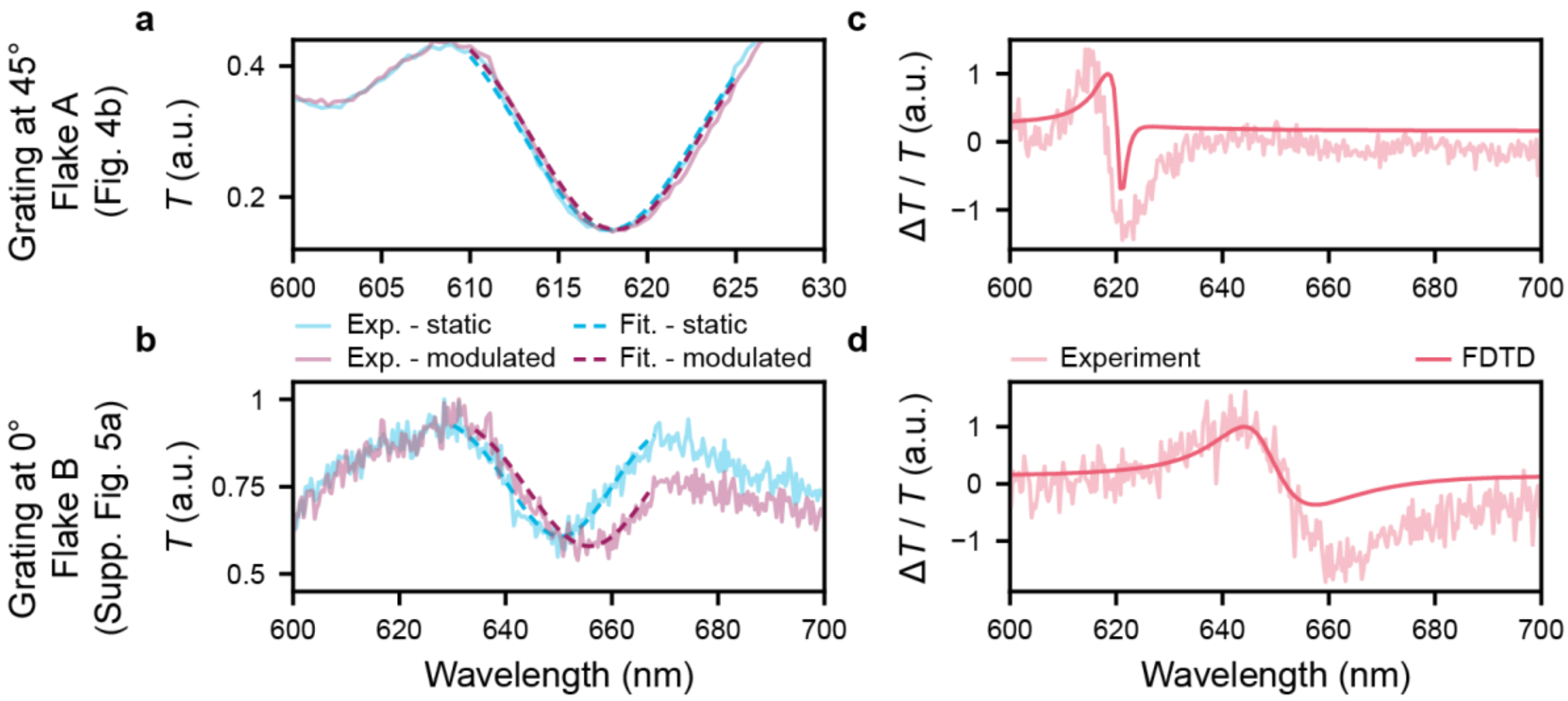

**Supplementary Figure 8.** Resonance shift in pump-probe measurements and FDTD modelling. **a.b** Transmitted probe spectrum before (blue continuous curves) and after modulation (purple continuous curves) for **a** the grating presented in the main text and **b** the grating presented in Supp. Fig. 5a, probing respectively the hybrid and plasmonic response. The resonant wavelength and FWHM are found by fitting a gaussian, here represented by the dashed curves. **c,d** Comparison between the relative change in transmittivity obtained experimentally (light red curve) and FDTD-simulated (red curve). FDTD simulation only assumes a 2% decrease in plasma frequency, $\Delta T/T$ is here shown in arbitrary units to demonstrate the agreement in the resonant wavelength shift between experiment and simulations.

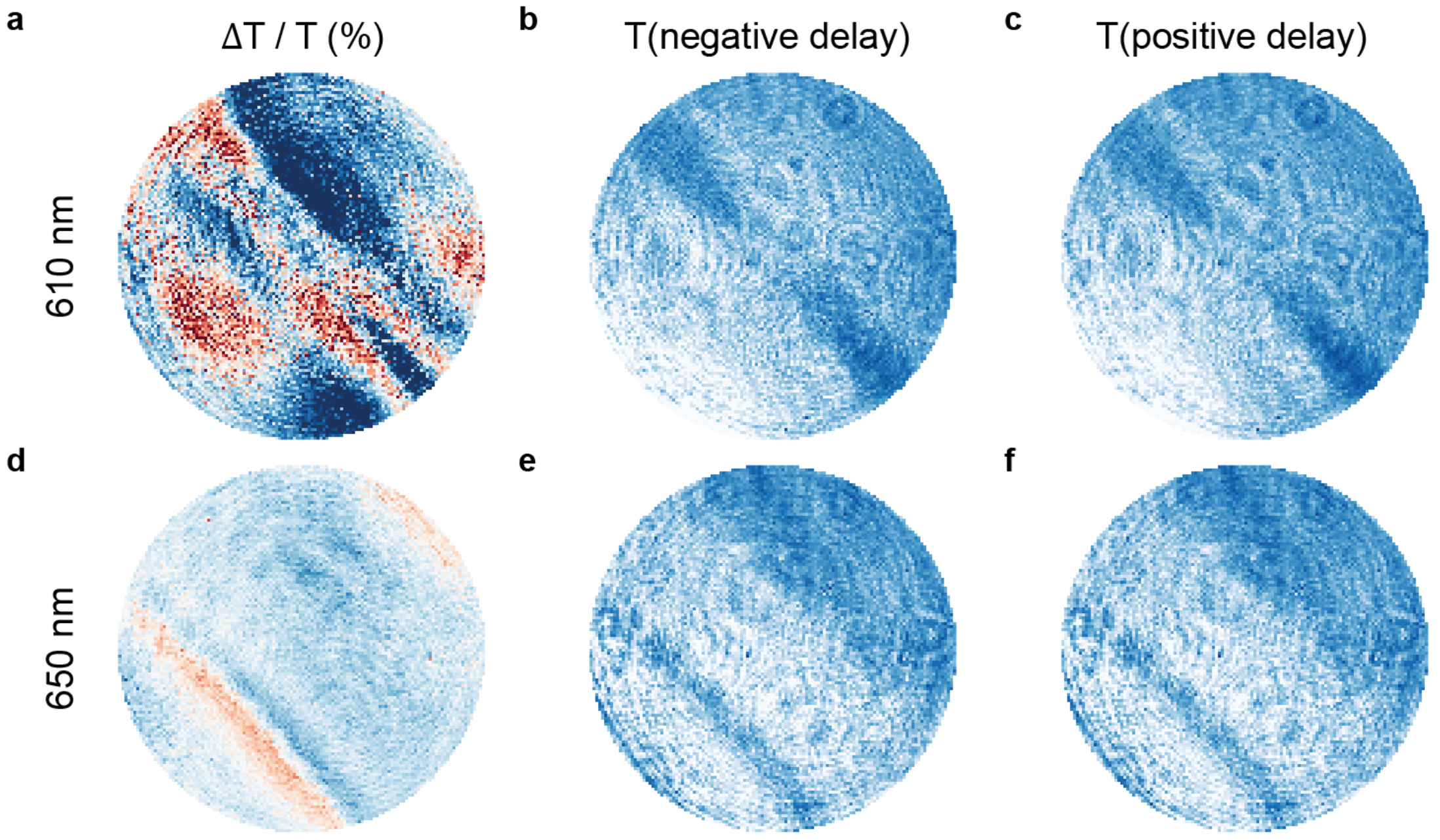

**Supplementary Figure 9.** Momentum-space pump-probe measurements. **a-c** Momentum-space image corresponding to the measurement presented in Fig. 4**d,f**: **a** relative change in transmission, **b** image at negative delay before the switching and **c** positive delay after the switching. **d-f** Momentum-space image corresponding to the measurement presented in Fig. 4**de,g**. Due to the short duration of the probe, the image is a convolution of the response over its bandwidth. In the case panels **d-f**, illumination of the probe near the edge of the grating (due to optical damage at the centre) leads to an asymmetric response in momentum space.

## FDTD simulations

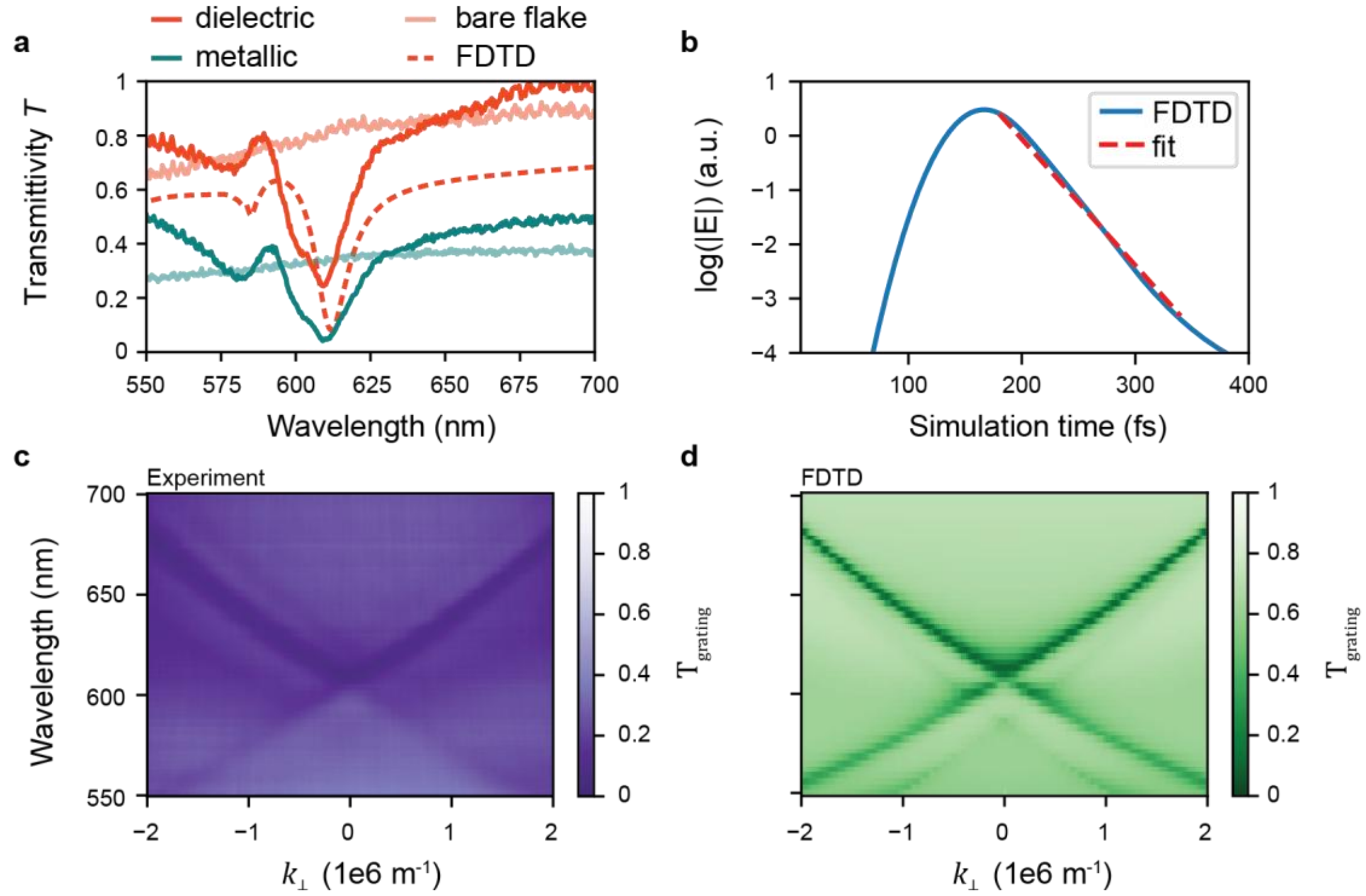

**Supplementary Figure 10.** FDTD simulation of the grating dispersion and comparison with experiment. **a** Linear transmission spectrum of the grating from FDTD (orange dashed line) superimposed with the experimentally measured transmission curves for dielectric (orange curves) and metallic (green-blue curves) incident polarisation. The partially-transparent curves correspond to the transmission from the bare flake next to the grating. **b** Measured exponential decay in time of the electric field 2 nm above the surface of the device for an incident probe at 595 nm wavelength (corresponding to Fig. 3**a**), giving a Q-factor of 69.7. **c** Experimentally measured dispersion curve corresponding to Fig. 3**c**. **d** FDTD-simulated dispersion curve.

## Nonlinear sensitivity of the hybrid modes: TE against TM

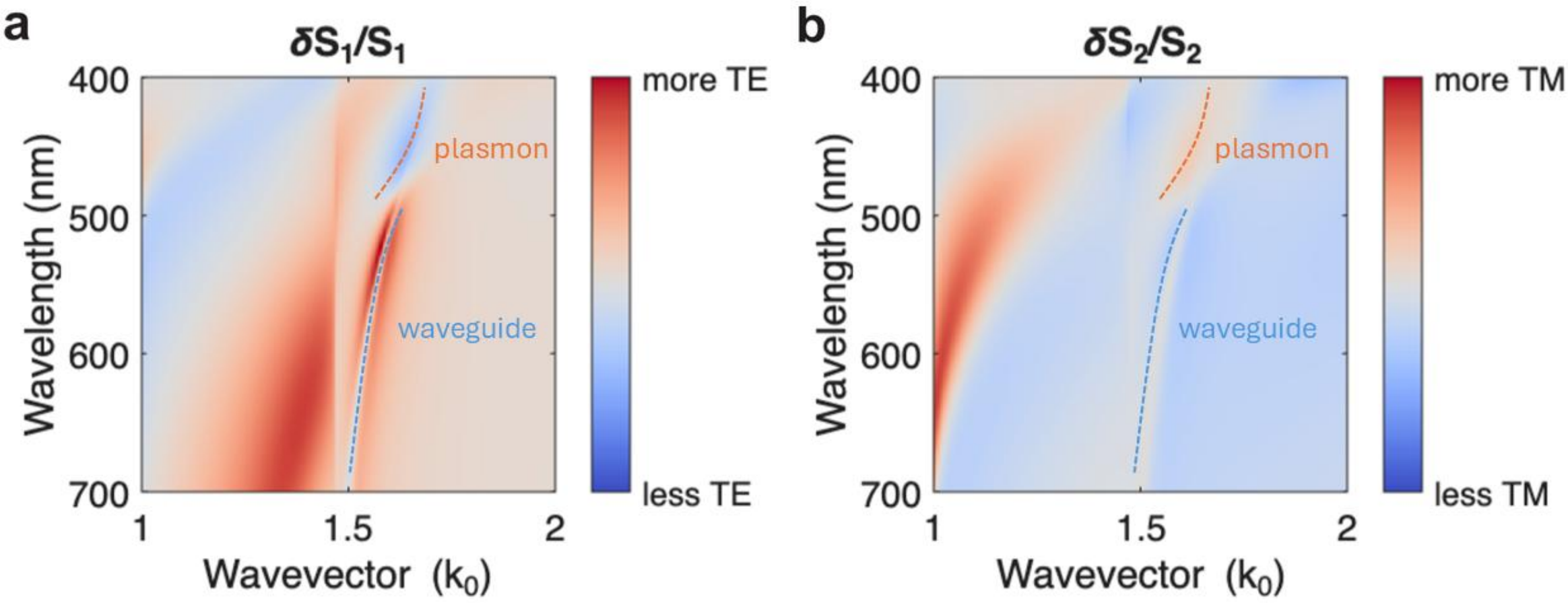

**Supplementary Figure 11.** TMM simulation of the hybrid modes under optical pumping. **a** Change in the TE nature of the mode $\delta S_1/S_1$ where $S_1$ is defined as $\left(r_{TE}^{(i)} - r_{TM}^{(i)}\right) / \left(r_{TE}^{(i)} + r_{TM}^{(i)}\right)$ and $r_{TE/TM}^{(i)}$ is the imaginary part of the reflection coefficient under TE or TM excitation. **b** Change in the TM nature of the modes $\delta S_2/S_2$ where $S_2$ is defined in the same way as $S_1$ for TM polarisation.